\documentclass[twocolumn,groupedaddressm,showkeys,byrevtex,amscd,amsmath,amssymb,verbatim]{revtex4}
\usepackage{graphicx}
\usepackage{bm} 
\usepackage[dvips]{hyperref}
\usepackage{textcomp}

\begin{document}

\title{Conceptual Design of a Micron-Scale Atomic Clock}
\author{Eric C. Hannah}
\email{eric.hannah@intel.com}
\affiliation{Intel Corporation}\author{Michael A. Brown}
\affiliation{Intel Corporation}\label{I1}
\date{\today}
\begin{abstract}
A theoretical proposal for reducing an entire atomic clock to micron
dimensions. A phosphorus or nitrogen atom is introduced into a fullerene cage. This endohedral
fullerene is then coated with an insulating shell and a number of them are deposited as a 
thin layer on a silicon chip. Next to this layer a GMR sensor is fabricated which is close to
the endohedral fullerenes. This GMR sensor measures oscillating 
magnetic fields on the order of micro-gauss from the
nuclear spins varying at the frequency of the hyperfine transition (413 MHz frequency). Given
the micron scale and simplicity of this system only a few transistors
are needed to control the waveforms and to perform digital clocking.
This new form of atomic clock exhibits extremely low
power (nano watts), high vibration and shock resistance, stability on the order of $10^{-9}$, and is compatible with
MEMS fabrication and chip integration. As GMR sensors continue to improve in sensitivity the stability of this
form of atomic clock will increase proportionately.

\end{abstract}
\pacs{PACS}
\maketitle

\section{Background}

Atomic clocks have been in existence since 1949. The
basic time keeping element is the hyperfine interaction between
outer electron(s) and the nuclear spin. Thus embodiments typically
use atomic hydrogen, the alkali metals, and ions that have a single
outer orbital electron remaining; however hyperfine splitting is
present in more general atomic and molecular species and can be
used for time keeping. The hyperfine interaction is robust against
perturbations from vibration or temperature since it only involves
the density and spin of the electron wave function at the nucleus. High quality
atomic clocks have time precision of better than 1 part in ${10}^{15}$.
Atomic clocks have been reduced to a few cubic millimeters in size.
There is a great deal of interest by DARPA and the electronics industry in reducing  the size, complexity, and power requirements
of atomic clocks. A precise and stable time base that can fit within
the packaging and power envelope of modern devices will greatly
increase the efficiency and robustness of mobile computing and sensor
elements. Spectral bandwidth allocation is fundamentally limited
by the accuracy and stability of the reference frequency used to
define the clock period. One of the fundamental limits to reducing
the size of electronic wireless sensors (motes) is the fact that
crystal oscillators cannot be shrunk beyond the current package size\cite{Culler}.
A MEMS-scale atomic clock could replace complex clock synchronization methods within
computers and improve arbitration protocols between chips by supplying
precise local clocks. 

\subsection{Atomic Clock Principles}

Currently atomic clocks operate by using vapors of Cs
or Rb coming off an oven at $\sim$80  $^{\circ}$C. This results
in about 1 atom per cubic micron. In some optical clocks buffer gas is used to slow the
collision rate of the atoms with the wall where spin flipping occurs
-- broadening the hyperfine line width. Each atom produces a magnetic
field due to the spin of the nucleus -- ${10}^{-11}$ gauss at 1 micron distance. When the atomic system
has been excited, the magnetic field oscillates at the hyperfine
transition frequency: 100 MHz to 20 GHz -- depending upon the atomic
species in use. The best current systems use stabilized lasers to
interact with the atomic system by means of optical transitions.
Thus they do not directly sense the nuclear magnetic field.
This method is costly in size and power.

\section{Endohedral Fullerenes}\label{XRef-Section-1061425}
\begin{figure}[h]
\begin{center}
\includegraphics[width=2.5in]{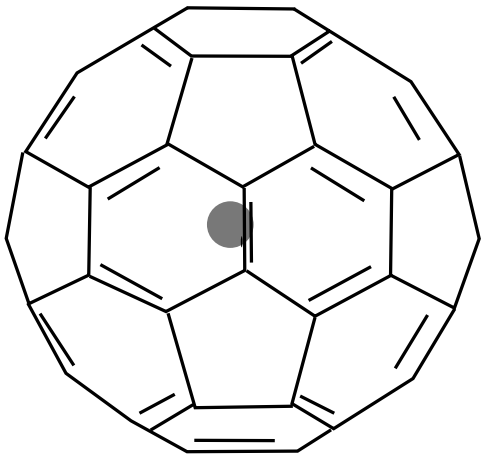}

\end{center}
\caption{Model of a Buckyball with an enclosed atom}
\label{XRef-FigureCaption-10614219}
\end{figure}

Buckyballs (C${}_{60}$ and C${}_{70}$, each about 1 nanometer in diameter) have been developed that
contain a single atomic passenger and even simple molecules\cite{Murphy,Weidinger,Shinohara,Saunders,Stevenson}.
Figure \ref{XRef-FigureCaption-10614219} shows a simple
model. For most of these endohedral fullerenes a charge transfer
of the enclosed atom to the fullerene cage occurs resulting in a
chemical bond and distorted structure. In the case of group
V atoms the trapped atom is confined by a harmonic-like potential
to the center of the cage. The atom is trapped within the covalent
bonds of the Buckyball with its outer electrons symmetrically repelled away from
the walls of the cage. In almost all respects the atom behaves
as a `free' (unbonded) atom, though spatially restricted to be within
the fullerene cage. The distributed $\pi$ bonding electrons also
act as an almost perfect Faraday cage. The inner atom is fully capable
of entering into magnetic interactions since the Buckyball cage
is spin neutral (it is desirable to use C{}\textsuperscript{12}
to remove nuclear spins within the cage walls). Two useful compounds
are N@C${}_{60}$ and P@C${}_{60}$, i.e., nitrogen and phosphorus
inside C${}_{60}$. Group V atoms are paramagnetic due
to their half-filled p-orbitals. Electron Spin Resonance experiments
have demonstrated that the trapped electrons for group V atoms have
very long spin relaxation times\cite{Harneit}.

Group V atoms can be implanted into fullerenes by simultaneous ion
bombardment and fullerene evaporation onto a target\cite{Harneit}.
The resulting endohedrals are themally\cite{Waiblinger} and chemically
stable\cite{Pietzak} at ambient conditions but do show disintegration
between 400K and 600K\cite{Waiblinger2001}. In low-energy implanted samples
only a small fraction are actually filled, typically 1 in 10,000.
It is possible to enrich and purify the fraction of filled molecules
by high-pressure chromatography\cite{Goedde}. Recent work has shown that nuclear implantation techniques (using 18 MeV ions) can convert almost 6\% of a thick fullerene  layer ($\sim$40 $\mu$m) into endohedral compounds \cite{Ray}. A $C_{60}$ Buckyball
with an atom trapped within has a nominal diameter of 1 nanometer.

\subsection{Endohedral Fullerenes Hyperfine Interactions}

Phosphorus comes in one stable isotope: ${}^{31}P(100\%)$, spin/parity
= $1/2$+.
Its electronic ground state, ${}^4S_\frac{3}{2}$, is split by the hyperfine
interaction. Nitrogen comes in two stable isotopes: ${}^{14}P( 99.6\%)
$, spin/parity = 1+ and ${}$${}^{15}N( 0.37\%) $, spin/parity =
$1/2$-.
Both isotopes have their ground state split by the hyperfine interaction.
For the sake of concreteness we focus on phosphorous. Nitrogen
may require isotopic separation for use in an atomic clock, or,
we could accept a double resonance mode of operation.

The contact part of the hyperfine interaction is
\begin{equation}
U = \gamma \hbar \mu_B {\left| \psi (0) \right|}^2 \mbox{\boldmath $I \cdot S$}
\end{equation}

where {\bfseries {\itshape I}} is the nuclear spin in units of $
\hbar $ ({\bfseries {\itshape S}} is the electon spin). Typical values of the hyperfine frequency are in the microwave range.

One important effect is the increase in electron density at the
nucleus due to confinement inside the fullerene. Recent experimental
research shows an enhanced electron capture decay rate in ${}^{7}\mathrm{Be}$
when it is encapsulated inside C${}_{60}$\cite{Ohtsuki} .
\begin{figure}[h]
\begin{center}

\includegraphics[width=2.5in]{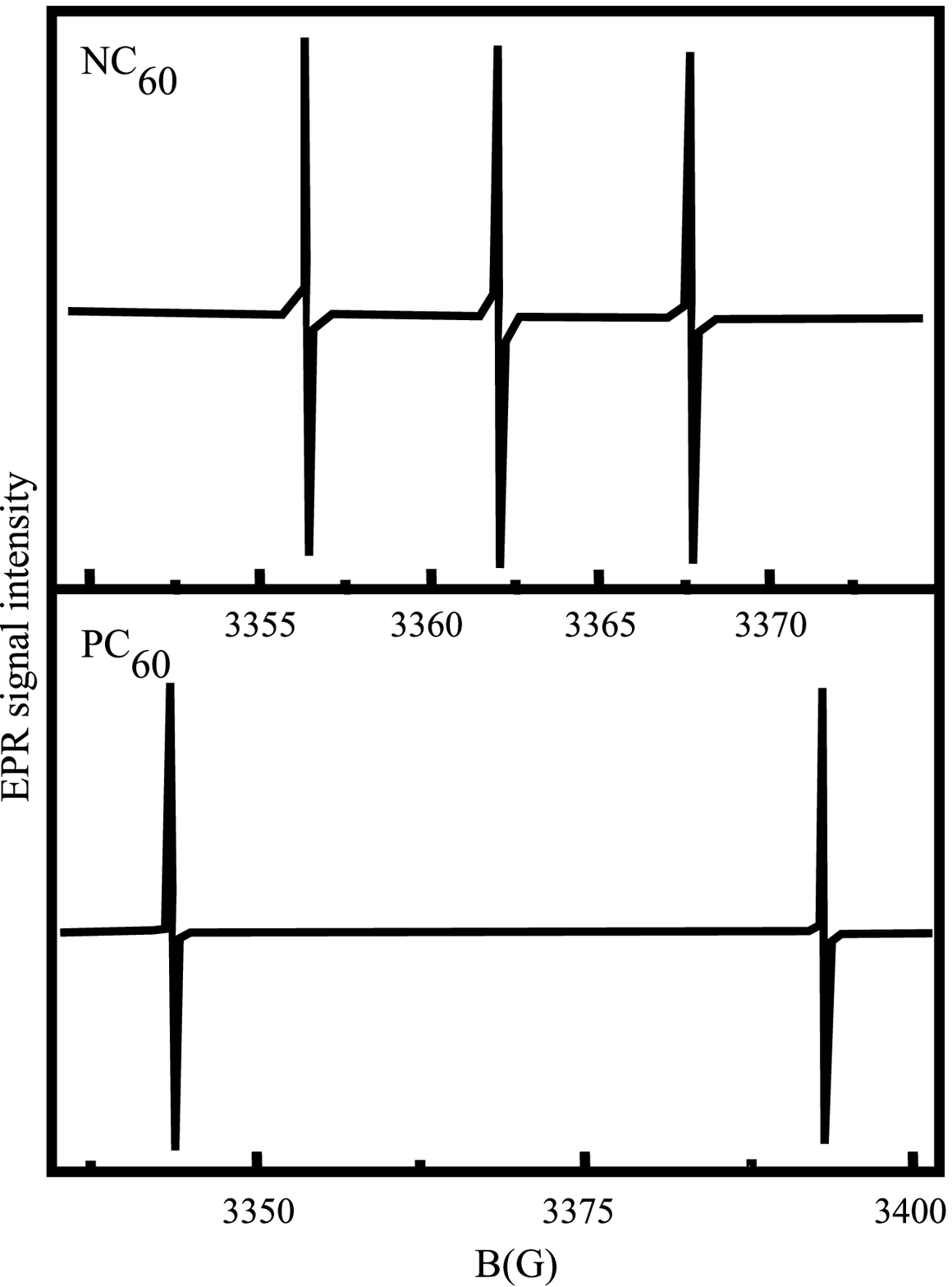}

\end{center}
\caption{EPR spectra, \cite{Waiblinger2001}}
\label{XRef-FigureCaption-10615488}
\end{figure}

Figure \ref{XRef-FigureCaption-10615488} shows EPR experimental
data\cite{Waiblinger2001} on the P@C${}_{60}$ ground state splitting
due to the hyperfine interaction. The doublet splitting ($\Delta$=49.2
G) in the lower panel is due to the hyperfine interaction of the
electron spin with the nuclear spin I=$1/2$
of ${}^{31}\mathrm{P}$. The EPR spin-magnetic field interaction
energy is
\begin{equation}
U = \pm 3 \mu _{B}B_{a}%
\label{XRef-Equation-112315628}
\end{equation}

This implies a hyperfine frequency of 413 MHz. This is a fairly
low hyperfine frequency but reflects that fact that the outer
electrons have finite angular momentum, consequently they have zero
spatial overlap at the nucleus. The hyperfine interaction in paramagnetic
atoms or ions is entirely determined by the admixture of electronic
{\itshape s}-orbitals from excited configurations which result from
atomic interactions such as electron repulsion and correlation\cite{Abragam}.

\subsection{Atomic Transitions}\label{XRef-Subsection-1031183916}
\begin{figure}[h]
\begin{center}
\includegraphics[width=2.5in]{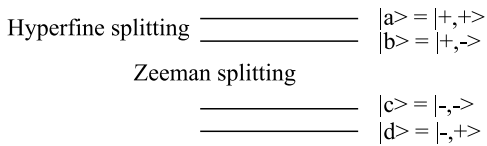}

\end{center}
\caption{Zeeman/hyperfine energy levels, label = $ | S_{electron}$, $I_{nuclear}\rangle$}
\label{XRef-FigureCaption-1031183928}
\end{figure}

Given that P@C${}_{60}$ is a paramagnetic atom the most straightforward
method for measuring the hyperfine energy is to polarize
the atom in a large static magnetic field. Figure \ref{XRef-FigureCaption-1031183928}
shows the atomic energy levels when the electron spin - magnetic
field interaction energy exceeds the hyperfine energy splitting.
In the absence of EPR pumping the vast majority of atomic systems
will be in the lowest energy state. The polarized outer electron
spins will define a spatial direction for the nuclear spin and will
define its energy transitions. Usually hyperfine transitions inside
paramagnetic atoms and ions have a poorly defined line width
-- due to the vast number of perturbations acting on the valence
electrons, which are directly relayed to the nucleus through the
hyperfine interaction. In the case of P@C${}_{60}$ the exceptionally
long electron spin relaxation times and the nearly ideal conditions
of isolation give us relief from these problems.

\subsection{Interactions Between Endohedral Fullerenes}

One issue in the use of a packed array of P@C${}_{60}$ molecules
for an atomic frequency standard is the perturbation nearby spins
apply to each other. Nearby electron spins will contribute varying
magnetic energy terms to the nuclear spin. As each nuclear spin will
see a slightly different electron spin neighborhood this will broaden
the hyperfine frequency of the ensemble. The magnitude of the shift for each 
electron-nucleus pair is
\begin{equation}
f_{\mathrm{nuclear \, shift}}=\pm 3 \frac{\mu _{0}\mu_{B}\mu_{n}}{4\pi  h r^{3}}
\end{equation}

Table \ref{XRef-TableTitle-2} shows this interaction for an electron-nuclear pair as a function of separation.

\begin{table}
\caption{Nuclear-Electron Perturbation Effects} 
\label{XRef-TableTitle-2} 
\begin{ruledtabular}

\begin{tabular}{lc}
Distance [nm] & Frequency Shift\\
\hline
1  & 10 KHz\\
10  & 10 Hz\\
100  & 10 mHz\\
1000 & 10 $\mu$Hz
\end{tabular}
\end{ruledtabular}
\end{table}

This implies that a close-packed solid of endohedral fullerenes
(diameter = 1nm) will exhibit a fractional frequency shift on the
order of ${10}^{-4} {\mathrm{N}}_{\mathrm{nearest} \, \mathrm{neighbors}}$.
In itself this is not a problem as this is a constant shift. The
problem stems from the variability of the packing. Randomness in
the distances to nearby endohedral fullerenes will modulate the
perturbation. This random perturbation will broaden the absorption
line of the system resulting in poor short-time frequency control.
Assuming stable packing the long term frequency control of the system
will still be exemplary. 

Since the perturbation term varies as $\frac{1}{r^{3}}$a simple
solution could be to dilute the endohedral fullerenes inside a neutral
carrier material. The issue now is to estimate the effect of random
spacing between endohedral fullerenes and the corresponding perturbation
effects so that the net line width of the system meets the short-time
frequency requirement. The main problem is the variance in the frequency.
Appendix \ref{XRef-AppendixSection-1010145454} calculates this effect
with the results in Table \ref{XRef-TableTitle-1010145529}.

\begin{table}
\caption{Density Effect on the Standard Deviation of the Frequency
Shift($R_{\mathrm{cutoff}}$ = 1nm)} \label{XRef-TableTitle-1010145529}
\begin{ruledtabular}

\begin{tabular}{lc}
Density of fullerenes [$m^{-3}$] & $\sigma _{\text{$\langle f_{\mathrm{nuclear} \,
\mathrm{shift}}\rangle $}}$ [Hz]\\
\hline
${10}^{17} $ & 5\\
${10}^{18} $ &  17\\
${10}^{19} $ & 55 \\
${10}^{20} $ & 172 \\
${10}^{21} $ & 545\\
${10}^{22} $ & 1,721\\
${10}^{23} $ & 5,384\\
${10}^{24} $ & 16,086\\
${10}^{25} $ & 39,288\\
${10}^{26} $ & 62,452
\end{tabular}
\end{ruledtabular}
\end{table}

The conclusion from this calculation is that the divergent
$\frac{1}{r^{3}}$ interaction amplifies the effect of statistical
fluctuations inside a homogenous mixture. While the effects of randomness in fullerene locations limit the ultimate precision of an atomic clock, for less demanding uses, e.g., crystal clock replacement, a simple layer of $\mathrm{C}_{60}$ doped with P by nuclear implantation techniques should prove an attractive option.

A random mixture of endohedral
fullerenes is not the best spatial organization for a frequency standard.
The best arrangement is a uniform grid with a few 10's of nm spacing
between occupancy sites.

\subsection{Glassification of Endohedral Fullerenes}

It is possible to separate each endohedral fullerene from its
neighbors by coating it with a glass shell. Silica gel, an inorganic
polymer, has a three-dimensional network and can easily be synthesized
via the sol-gel route. Fullerenes cannot be incorporated into sol-gel
glasses homogeneously due to low solubility. This problem can be
overcome by functionalization of the fullerenes with such groups
as will form some kind of bond (hydrogen, van der Waals, or covalent)
with the growing silica network\cite{Patwardhan}.\ \ The synthesis
of silica-fullerene hybrid materials is done in the following manner.
Silica precursor is mixed with alcohol and water. Acid is added
as a catalyst. Functionalized fullerene is added to this mixture
directly or by dissolving in toluene. The mixture is allowed to
gel over a period of time at elevated temperature. Coatings of single
fullerene molecules can be accomplished by performing the above
steps in a micro-fluidic mixing chamber where the timed release
of materials results in small globules of glass-encased single fullerene.

Fullerene functionalization consists of covalently bonding side
groups mainly consisting of nitrogen or Si-O-R (R= Me, Et) or even
hydroxyl groups. The molecular carbon allotrope readily adds nucleophiles
and carbenes and participates as the electron-deficient dienophile
component in many thermal cycloaddition reactions such as the Diels-Alder
addition. In all mono-adducts formed by 1,2-addition the fullerene
preserves the favorable $\pi$-electron system of C${}_{60}$\cite{Diederich}, however, we can expect a loss of the full symmetry in Buckyball shielding -- the net effect on the hyperfine energy of the system needs to be determined by experiment.
\begin{figure}[h]
\begin{center}

\includegraphics[width=2.5in]{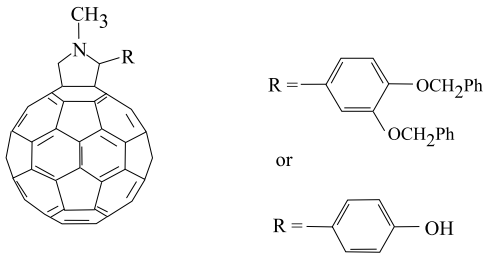}

\end{center}
\caption{Functionalization of a fullerene for sol-gel coating, \cite{Patwardhan}}

\end{figure}
\begin{figure}[h]
\begin{center}

\includegraphics[width=2.5in]{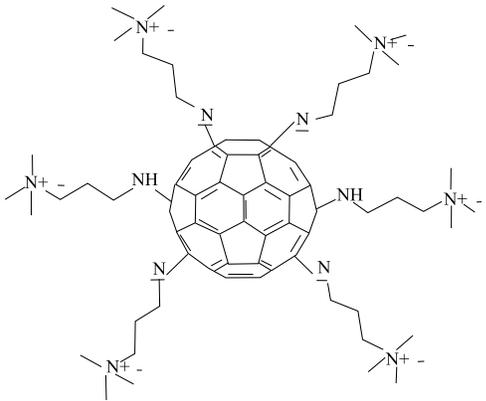}

\end{center}
\caption{Functionalized fullerene inside a sol-gel network, \cite{Patwardhan}}

\end{figure}

So we see that a good arrangement for the endohedral fullerenes is to
coat them with $\sim$ 20nm diameter glass coating. This
separates the electron-nuclear spins to the point where a narrow
line width is possible. It should be noted that wrapping polymers around fullerenes
is also a separation option.

Another embodiment which avoids the possible problems of functionalization is to create simple fullerenes and then implant them at low energy into an insulating matrix. We can also use patterning (e.g., 2D interference patterns from hard or soft UV light) to create openings in a photoresist layer. Next we implant a low flux of fullerenes so as to only deposit at most one fullerene per opening. Follow by a thin layer of an insulating material and so on. Finally the photoresist layer with the extra fullerenes is burned off by an oxygen plasma. There are many techniques used in electronics that can create well-separated fullerene matrices.

\subsection{Magnetic Parameters for Encapsulated Endohedral Fullerenes}

If we assume a uniform sample of $\sim$ 20nm diameter
glass spheres with a single fullerene inside we predict an internal
magnetic field from the nuclear spins of
\begin{equation}
B_{\mathrm{nucleus}} \cong  \frac{\mu _{0}}{4\pi }\mu _{n}N_{\mathrm{density}}
= 10^{-10} [\mathrm{tesla}] = 1 \mu \, \mathrm{gauss}.
\end{equation}

At the same time the internal magnetic field from the electron spins is
\begin{equation}
B_{\mathrm{electronic}} \cong  \frac{3 \mu _{0}}{4\pi }\mu _{B}N_{\mathrm{density}}
= 3 \times 10^{-7} [\mathrm{tesla}] = 3 \, \mathrm{milli} \mathrm{gauss}.
\end{equation}

Because the source of the polarizing field for the nuclear spin
is the electron spin coupled through the hyperfine interaction we
need to re-express this in terms of the effective magnetic field
being applied to the nucleus. Translating the 413
Mhz hyperfine frequency into a nuclear Zeeman splitting gives an
equivalent magnetic field of 27 tesla. For the coupling term we have
\begin{multline}
S_{\mathrm{electron}} =  3/2,\\
U_{\mathrm{Hyperfine}} = 4.13 \times 10^8 \mathrm{[Hz]}\; h \; \mbox{\boldmath $I \cdot S$} = 2.75 \times 10^{-25} \; \mbox{\boldmath $I \cdot S$} \mathrm{[Joules],}\\
\mbox{\boldmath $B_{\mathrm{effective}}$} = \frac{U_{\mathrm{Hyperfine}}}{\mu_N}
\mbox{\boldmath  $S$}
= \frac{54 \mathrm{[Tesla]}\mbox{\boldmath  $S$}}{6} = 9 \mbox{\boldmath  $S$}.
\label{XRef-Equation-666}
\end{multline}

It should be noted that the interaction energy represented by this pseudo magnetic field is very precisely defined -- it is the hyperfine energy. This expression for the interaction will allow us to directly employ the Bloch equations below to predict electron spin precession perturbations on nuclear polarization.

One of the design challenges for the atomic clock is to "pin" the electron spins
to minimize their precession which causes interference with the nuclear signal. Also, electron spins are very susceptible to environmental noise so we need to ensure that most of the excitation energy for the hyperfine defined nuclear spin flips come from an external AC drive field directly -- not through the precessing of the electron spins.

\subsubsection{The Bloch Equations}

In our approach there are three spin mechanisms at work.
The first mechanism is NMR where the DC polarization field comes from the spin polarized electrons via the hyperfine interaction and nuclear spin flips are driven by magnetic resonance with an external AC magnetic field. By design this mechanism is on resonance. Bloch's equations describe this situation accurately as there are a large number of noninteracting spins undergoing precession. The second mechanism is induced precession in the outer electron spin with an external DC magnetic field and the same AC magnetic field as for the NMR mechanism.  With a large Zeeman splitting this precession is far off resonance and the
electron net spin will only precess, never flip. We need to reduce this signal
to a small value that won't interfere with the nuclear hyperfine signal. The third
mechanism is nuclear spin precession driven by the precessing outer electron spin through the hyperfine interaction. Bloch's equation, using the equivalent magnetic field of the hyperfine interaction, works here as well. We need to ensure that this
mechanism is small compared to the conventional magentic excitation. In all three cases we can use the Bloch approach.

Bloch's phenomenlogical equation is an adequate description for both the electron  and nuclear spin
\begin{multline}
\frac{d \text{\boldmath $\mathrm{M}$}}{d t}= \gamma  \text{\boldmath
$\mathrm{M}$}\text{\boldmath $\wedge $}\text{\boldmath $\mathrm{H}$}
\text{\boldmath $\mathrm{-}$} \frac{M_{x}\text{\boldmath $\mathrm{i}$}'+M_{y}\text{\boldmath
$\mathrm{j}$}'}{T_{2}}-\frac{M_{z}-M_{0}}{T_{1}}\text{\boldmath
$\mathrm{k}$}',\\
\end{multline}

where \mbox{\boldmath $i', j', k'$} are the unit vectors of the laboratory frame of reference, $T_1$ and $T_2$ are phenomenological relaxation times, and the various $M's$ are the magnetization vector components.

In the prototypical magnetic resonance experiment a large DC magnetic field, ${\text{\boldmath
$\mathrm{H}$}}_{0}$, is applied along the $\text{\boldmath $\mathrm{k}$}'$direction. An rf magnetic field, ${\text{\boldmath
$\mathrm{H}$}}_{1}$, is applied along either the $\text{\boldmath
$\mathrm{i}$}'$ or $\text{\boldmath $\mathrm{j}$}'$direction.

After a series of calculations Abragam\cite{Abragamp46} derives the
following time-dependent magnetization in the laboratory reference
frame
\begin{multline}
M_{y} \cong  \gamma  H_{1} M_{0} f_{T_{2}}( \Delta \omega )  \cos
\omega  t ,\\
f_{T_{2}}( \Delta \omega )  = \frac{T_{2}}{\pi }\frac{1}{1+{\Delta
\omega }^{2}T_{2}^{2}},\\
\Delta \omega  =\omega  - \omega _{\begin{array}{l}
 0,
\end{array}}\\
\gamma = \mathrm{gyromagnetic} \, \mathrm{ratio},\\
M_{0} = \mathrm{the} \, \mathrm{magnetic} \, \mathrm{moment} \, \mathrm{of} \,
\mathrm{the} \, \mathrm{sample},\\
\omega _{0}= \gamma  H_{0},\\
\omega \ \ =\ \ \mathrm{rf} \, \mathrm{oscillation} \,\mathrm{frequency},\\
H_{1}= \mathrm{magnitude} \, \mathrm{of}\, \mathrm{rf}\, 
\mathrm{field}\\
\label{XRef-Equation-112418355}
\end{multline}

This assumes negligible saturation, that is $\gamma ^{2}$$H_{1}^{2}$$T_{1}
T_{2} \ll  1$. For the three spin mechanisms involved we replicate the Bloch approach,
using the nuclear parameters in the first case, then the electronic parameters, and finally the hyperfine driven precession with $H_1$ now coming from the electron spin through the hyperfine effective magnetic field.

Twamley \cite{Twamley}
reports that the electronic relaxation times for Group V endohedral
fullerenes are $T_{1} \sim 1s$ at $T \sim 7^{o}K$, while
$T_{2} \sim 20\mu s$ for all temperatures. He also states that nuclear
relaxation times should be several orders of magnitude longer than
the electronic relaxation times. We will assume 1 second applies
for the nuclear relaxation parameters at room temperature.

For the pure nuclear NMR spin mechanism this implies that
$H_{1} \ll  {10}^{-8} \mathrm{tesla}$, or 100 micro-gauss. Equation
\ref{XRef-Equation-112418355} implies a resonance when the applied
rf frequency equals the hyperfine transition frequency. The line
width at half maximum is $\frac{1}{T_{2}}$.
Given the other parameters we predict an NMR magnetic signal on resonance
of about 1\% of the applied rf magnetic field amplitude.

For the electron precession effect we use the above equation in the regime where the precession frequency is far above the rf frequency (for the precession frequency to be much larger than 413 MHz the DC magnetic field must be much larger than 49 gauss).
\begin{multline}
M_{y} \cong  \gamma  H_{1} M_{0} f_{T_{2}}( \Delta \omega )  \cos
\omega  t ,\\
\Delta \omega  \cong -\omega_0 = -\gamma H_0, \\
f_{T_{2}}( \Delta \omega )  \cong \frac{T_{2}}{\pi }\frac{1}{1+{\Delta
\omega }^{2}T_{2}^{2}} \cong \frac{1}{\pi \gamma^2 H_0^2 T_2}\\
\label{XRef-Equation-112418311}
\end{multline}

From (\ref{XRef-Equation-112418311}) we predict an electron spin precession magnetic signal of
\begin{multline}
B_y^{\mathrm{electron}} =  \gamma  H_{1} \frac{3 \mu _{0}}{4\pi }\mu _{B}N_{\mathrm{density}}\frac{1}{\pi \gamma^2 H_0^2 T_2}  \cos
\omega  t , \\
B_y^{\mathrm{electron}} =  H_{1} \frac{3 \mu _{0}}{4\pi }\mu _{B}N_{\mathrm{density}}\frac{1}{\pi \gamma H_0^2 T_2}  \cos
\omega  t ,  \\
\gamma \cong \frac{3 \mu_{\mathrm{B}}}{\hbar} , \\
\mathrm{Hence} \\
B_y^{\mathrm{electron}} \cong  H_{1} \frac{\mu _{0}}{4 \pi^2 } \hbar N_{\mathrm{density}}\frac{1}{H_0^2 T_2^{\mathrm{electron}}}  \cos
\omega  t ,
\label{XRef-Equation-112418312}
\end{multline}

We note that this signal drops as one over the DC magnetic field squared. For a 100 gauss $H_0$ field and $H_1 \sim 10^{-8}$ tesla we predict an electron precession magnetic field of $\sim 10^{-18}$ tesla. This is clearly not a problem against the 
$~10^{-10}$ tesla NMR output signal.

For our third spin mechanism (the hyperfine pseudo magnetic field) we see that an electron spin precession signal of $\sim 10^{-18}$ tesla versus the $\sim 10^{-7}$ tesla total electron magnetic field in the fullerene sample implies the fractional electron spin moment precessing about the DC magnetic moment is around $3 \times 10^{-12}$. From Equation \ref{XRef-Equation-666} we predict that perturbation at the nucleus is $~ 10^{-11}$ tesla. This is much smaller than the NMR rf field strength.

Thus to establish a strong and stable electron spin system within which to monitor nuclear spin transitions we need an external DC magnetic field of at least
a few hundred gauss. Pure iron has a magnetization of 1700 gauss at room temperature. Thus a simple ferromagnet can provide the polarization field.

\section{Magnetic Sources}

While a coil of wire is a convenient source of uniform magnetic
fields when dealing with macroscopic dimensions, this is not true
on the micron scale. Fabrication of coils in IC's demands the use
of vias and complex materials fabrication. A better approach is
to use Maxwell's equations and derive the magnetic field from the
displacement current inside a capacitor.

\subsection{Dimensional Analysis}

To show that the use of capacitors and typical CMOS operating voltages
can supply us with an adequate rf magnetic field we first perform
a crude calculation of the rough scale of the phenomena. We assume
a micron scale for the critical dimensions,
\[
\epsilon _{0} \sim {10}^{-11}F/\mathrm{m} \mathrm{,} \, {\mathrm{\mu }}_{0} \sim {10}^{-6}
H/\mathrm{m} \mathrm{,} \, d \sim 1 \mu \mathrm{m} \mathrm{=} {\mathrm{10}}^{-6}\mathrm{m}
\]

This implies some characteristic sizes for capacitor and inductors
\[
C \sim {10}^{-17}\mathrm{F}\mathrm{,} \, \mathrm{L} \sim {\mathrm{10}}^{-12}\mathrm{H}
\]

We estimate the drive current a micron-size inductor
needs to create a magnetic field sufficient to saturate the spin
ensemble, ${10}^{-8}$tesla.
\begin{multline*}
\Phi  = B A= L I \ \ = {10}^{-12}\mathrm{H} \mathrm{I}\mathrm{,}
\\
\mathrm{A} \sim {\mathrm{10}}^{-12}m^{2}\mathrm{,}\mathrm{}\\
\mathrm{B} \mathrm{=} {\mathrm{10}}^{-8} \mathrm{tesla}.
\end{multline*}

Hence I $\sim$ ${10}^{-8}$amps. We now want to determined
the drive voltage a typical micron scale capacitor must have to
create this amount of current through the displacement current given
by Maxwell's equations.
\begin{multline*}
Q = C V\mathrm{,}\\
\mathrm{I} \mathrm{=} {\mathrm{10}}^{-8}\mathrm{amps} = \dot{\mathrm{Q}}
= C \dot{V} = C \omega  \tilde{V}\mathrm{,}\\
\mathrm{\omega } \sim {\mathrm{10}}^{9}\mathrm{rad}/\sec 
\end{multline*}

Hence $\tilde{\mathrm{V}} \sim 1 \mathrm{volt}$. Thus applying an rf
voltage of around 1 volt to a typical micron scale capacitor will
generate the needed ${10}^{-8}$tesla magnetic field needed to spin
polarize the endohedral fullerence atomic standard.

Given that this calculation is simply dimensional analysis we conclude
that choices in geometrical ratios and materials constants, i.e.,\ \ permittivity
and permeability, can change these magnitudes by several orders
of magnitude -- in both directions.

\subsubsection{Example Magnetic Driver}

A concrete example of the above magnetic generator is given below.

\begin{figure}[h]
\begin{center}

\includegraphics[width=2.5in]{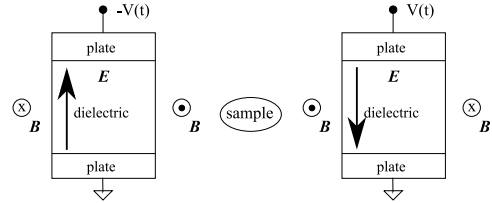}

\end{center}
\caption{Balanced capacitors magnetic source}

\end{figure}

Here two parallel plate capacitors (the long dimension of the plates
into the plane of the paper) generate an AC magnetic field between
them. The source of the magnetic field is the displacement current
inside the dielectric layers, i.e., the time varying electric field.
The scale of the external field's spatial extent is about the thickness
of the dielectric spacers. Using two capacitors driven out of phase
generates a low gradient magnetic field between them. This is a
two-dimensional version of Helmholtz coils.

\section{GMR Sensors}

The GMR effect takes place in heterogeneous magnetic systems
with two or more ferromagnetic components and at least one nonmagnetic
component\cite{Tsang}. An example is the trilayer permalloy/copper/permalloy
system. The GMR coefficient for a multilayer system is defined as the fractional
resistance change between parallel and antiparallel alignment of
the adjacent layers. This coefficient can be as high as 10\% for
trilayer systems and more than 20\% for multilayer systems.
\begin{figure}[h]
\begin{center}

\includegraphics[width=2.5in]{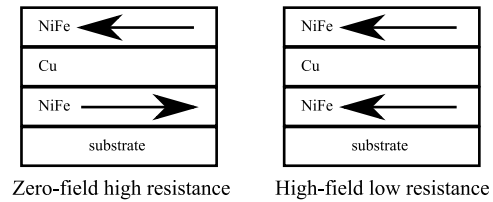}

\end{center}
\caption{The Giant Magnetoresistance effect is due to the large
difference in electrical resistance between two magnetic states
of a metallic multilayer film}

\end{figure}

 The occurrence of the GMR effect depends on the ability of the
applied magnetic to switch the relative orientation of the magnetic
moments back and forth between the parallel and antiparallel states.
In some multilayers a quantum-mechanical interlayer exchange coupling
across Cu or another paramagnetic metal causes a zero-field antiparallel
alignment which can be overcome by a high applied field. The magnitude
of the GMR effect can be surprisingly large, up to 80\%. However,
the fields needed to saturate Co/Cu multilayers are too large for
sensor applications. Other multilayers are designed to have an antiparallel
state in a limited applied field range by alternating ferromagnetic
layers (Co and Fe layers instead of two NiFe layers) with different
intrinsic switching fields. Thus the behavior of a multi-layer GMR
stack can be tuned to varying DC magnetic bias conditions.

Another interesting development is the ballistic magnetoresistance
effect, BMR, in ferromagnetic nanocontacts. The BMR effect arises
from nonadiabatic spin scattering across very narrow, $\sim$ atomic
scale, magnetic domain walls trapped at nano-sized constrictions.
In one study, the observation of a remarkably large room-temperature
BMR effect in Ni nanocontacts was reported. The observed BMR values
are in excess of 3000\% and are achieved at low switching fields,
less than a few hundred oersteds.\cite{Chopra}
\begin{figure}[h]
\begin{center}

\includegraphics[width=2.5in]{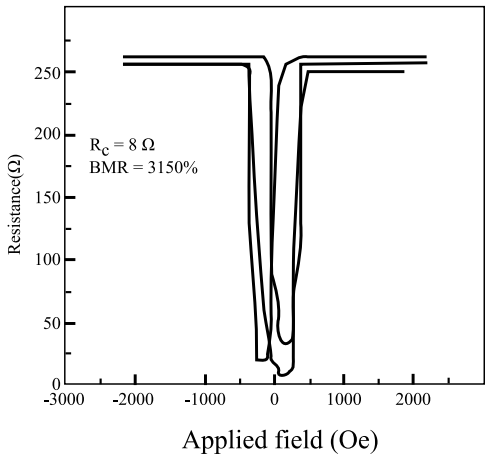}

\end{center}
\caption{Successive BMR loops from a Ni nanocontact showing 3150\% BMR, \cite{Chopra}}

\end{figure}

TMR (tunneling magnetoresistance) and CMR (colossal magnetoresistance) sensors are also devices that can sense magnetic state with increased sensitivity and lower power.

GMR read heads for hard disks operate at gigahertz frequencies.
This operation frequency is near the ferromagnetic resonance of
these devices. One study of these devices measured a resonance frequency
of 3.60 GHz\cite{Russek}. A recent experimental report shows a GMR
sensor sensitivity at room temperature of 350 pT/${\mathrm{Hz}}^{1/2}$\cite{Pennetier}.

\subsection{GMR Directional Sensitivity}

For atomic clock application it is desirable that the GMR sensor
reject the magnetic field coming from the both the strong DC polarizing
magnetic and the AC driving field that induces NMR resonance. The
DC field is easily rejected if it is directed perpendicular to the
multi-layer planes of the GMR sensor stack -- the various magnetic
layers only have substantial magnetic polarization in their plane.
The driving AC magnetic field is more difficult to reject for multi-layer
GMR stacks as it will be aligned along the planes of the stack,
just as is the atomic clock magnetic signal.\ \ If we choose to
detect the NMR signal that is perpendicular to the driving field
we have a chance to selective shield the GMR sensor from the AC
driving field direction by the use of a slitted Faraday shield.
\begin{figure}[h]
\begin{center}
\includegraphics[width=2.5in]{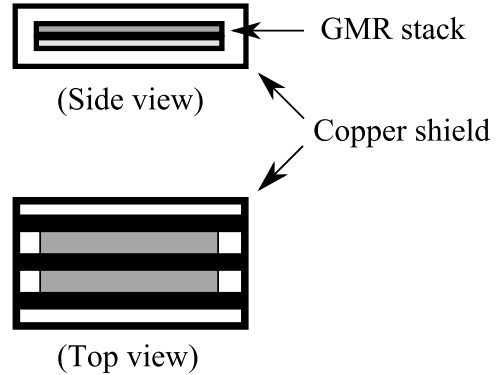}

\end{center}
\caption{Faraday shielding for GMR directional selectivity}

\end{figure}

The use of a slitted copper enclosure greatly reduces the GMR sensitivity
to magnetic field aligned perpendicular to the direction of the
slits while perserving sensitivity to parallel magnetic fields.
This holds only for high frequency magnetic fields and is due to
the induced eddy currents that counter the external applied field.

\section{Apparatus}

We now can put together a conceptual design for the atomic clock.
\begin{figure}[h]
\begin{center}
\includegraphics[width=2.5in]{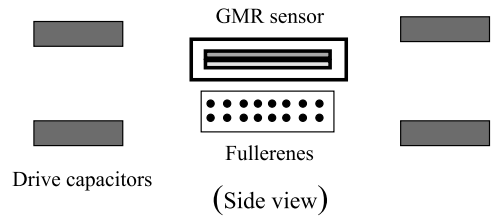}

\end{center}
\caption{Conceptual atomic clock design}

\end{figure}

\subsection{Atomic Clock Design}

Putting together all the ingredients discussed above we have a micron-scale
scheme that takes a drive voltage near the hyperfine resonance (driving
the capacitors) and produces a small ($\sim$ 1\% of the
drive field) signal that is detected and amplifed by the GMR sensor into a reasonable
voltage. Exact device values are highly dependent upon materials
and fabrication and won't be calculated here. In terms of basic
physics we expect a fullerene signal on resonance (413 MHz with
a line width $ \leqslant $ 1 Hertz) on the order of ${10}^{-10}$tesla
and GMR sensors at room temperature have achieved a sensitivity
of 350 pT/${\mathrm{Hz}}^{1/2}$. Thus in a 1 Hertz filtered loop
we should detect an atomic clock resonance with S/N $>$ 1.

\subsection{Electronics}

\begin{figure}[h]
\begin{center}

\includegraphics[width=2.5in]{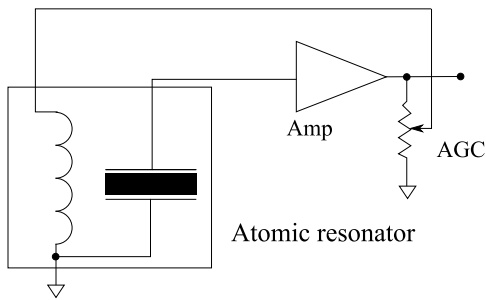}

\end{center}
\caption{Simple clock circuit}

\end{figure}

The above figure shows a simple oscillator design that will provide
an ouput sine wave locked to the line frequency of the fullerene
clock standard. The AGC circuit is designed to provide stable oscillations
just below saturation. Starting from Power On, the intrinsic noise
in the circuit will provide a nanovolt or so in the 1 Hz resonance
bandwidth. If the amplifier ultimately provides a drive signal of around
1 volt with a mild amount of net circuit gain, we should have stable
full-power operation in $\sim$ 100 system time constants,
$\sim$ 100 x 1 sec (1/bandwidth). Faster turn on can
be provided by strobing the input of the resonator with a pulse
providing substantial fourier components near the resonance.

\subsection{Clock Scalability}

The simple scheme we've discussed gives us a micron-scale atomic
clock with ${10}^{-9}$accuracy and a power dissipation of a nanowatt (10 nW capacitive drive but we can use resonant
circuits to store the energy). This will likely be adequate for
many mobile/sensor net applications but not adequate for more demanding
situations. What can be done?

First, as GMR sensors improve (BMR, etc.), we can use more diluted
fullerene stacks to gain a sharper line by a cubic factor in separation
as we lose an equal amout of magnetic signal. A nanoscale-precise
placing of fullerenes would give us a very well determined perturbation
situation that can be exploited for accuracy. In the limit of true
nanotechnology the ultimate clock is a single fullerene with considerable
shielding. This should be competitive with very good atomic clocks
of vastly more volume.

\appendix

\section{Variance of Fullerene Electron-Nuclear Interaction}
\label{XRef-AppendixSection-1010145454}

The idealized variation in frequency shift due to nearby endohedral
fullerenes

$n = $ density of fullerenes,


\begin{eqnarray}
\langle N \rangle & = & \mbox{expected number of neighbors within} \, R \\
& = & \int _{0} ^R n 4 \pi r^2 d r ,
\end{eqnarray}

\begin{equation}
\left \langle N \right \rangle = \frac{4 \pi}{3} R^3 n .
\end{equation}

Therefore

\begin{equation}
R_{\langle N \rangle = 1} = \sqrt[3]{\frac{3}{4 \pi n}} .
\end{equation}

The expectation value for the nuclear frequency shift

\begin{equation}
\langle f_{\mbox{nuclear shift}} \rangle = 3 \frac{\mu_0 \mu_B \mu_n}{h} \left \langle \frac{1}{r^3} \right \rangle ,
\end{equation}

\begin{equation}
\left \langle \frac{1}{r^3} \right \rangle = \int _0 ^R n \frac{1}{r^3} 4 \pi r^2 d r = 4 \pi n \int _{R_{\mathrm{cutoff}}} ^{R_{\langle N \rangle = 1}} \frac{1}{r} d r ,
\end{equation}

\begin{equation}
\left \langle \frac{1}{r^3} \right \rangle = 4 \pi n \left [ \log(R_{\langle N \rangle = 1}) - \log(R_{\mathrm{cutoff}}) \right ] \equiv k_{\mathrm{mean}},
\end{equation}

\begin{equation}
\sigma ^2 _{(\frac{1}{r^3})} = \mathrm{variance} = 4 \pi n \int _{R_{\mathrm{cutoff}}} ^{R_{\langle N \rangle = 1}} (\frac{1}{r^3} - k_{\mathrm{mean}})^2 r^2 d r
\end{equation}

\begin{eqnarray}
\sigma ^2 _{(\frac{1}{r^3})} & = & \frac{4 \pi n}{3} ( -\frac{1}{R_{\langle N \rangle = 1}^3} + k^2_{\mathrm{mean}} R_{\langle N \rangle = 1}^3 + \frac{1}{R_{\mathrm{cutoff}}^3} \\
& & - k_{\mathrm{mean}}^2 R_{\mathrm{cutoff}}^3\nonumber - 6 k_{\mathrm{mean}} \log[R_{\langle N \rangle = 1}]\\& & + 6 k_{\mathrm{mean}} \log[R_{\mathrm{cutoff}}])  .
\end{eqnarray}


\end{document}